# Surface domain engineering in congruent lithium niobate single crystals: A route to submicron periodic poling

A. C. Busacca, C. L. Sones, V. Apostolopoulos, R. W. Eason, and S. Mailis[a]
*Optoelectronics Research Centre University of Southampton, Southampton SO17 1BJ, United Kingdom*

()

We describe a technique for surface domain engineering in congruent lithium niobate single crystals. The method is based on conventional electric-field poling, but involves an intentional overpoling step that inverts all the material apart from a thin surface region directly below the patterned photoresist. The surface poled structures show good domain uniformity, and the technique has so far been applied to produce domain periods as small as ∼1 $\mu$m. The technique is fully compatible with nonlinear optical integrated devices based on waveguide structures. []

Domain engineering in ferroelectric crystals such as $LiNbO_3$ and $LiTaO_3$ is an increasingly important and ever more versatile technique for applications in areas as diverse as harmonic generation and parametric processes,[1] electro-optic Bragg gratings,[2] and piezoelectric microactuated devices.[3] Over the past decade or so, highly efficient quasi-phase-matched nonlinear interactions have been achieved via precise periodic domain inversion in z-cut crystal samples, using periods for example of the order of a few $\mu$m for near-infrared to blue or near-UV harmonic generation.[4,5] Research on periodically poled lithium niobate, PPLN (and to a lesser extent lithium tantalate), continues to generate considerable interest from the fundamental viewpoint of materials research through to the fabrication of practical nonlinear optical and electro-optical devices. PPLN with periods for standard conversion wavelengths is now commercially available from several sources.

Fabrication of periodically poled materials with arbitrarily small values of period, particularly at submicron scales, remains an elusive goal however. The high coercive field, $E_c$, required for domain inversion in congruent $LiNbO_3$ ($E_c \sim 220$ kV cm$^{-1}$), together with inherent non-uniformities and defects that are always present in commercially available materials, restricts the applicability of the standard electric-field poling technique to periods on the order of >4–5 $\mu$m in samples of thicknesses ∼500 $\mu$m. It is not an easy task to routinely fabricate high-quality PPLN and, in many cases, the crystal must be polished down to thicknesses on the order of 100–150 $\mu$m to achieve finer periods than this.[6,7]

Two approaches to overcome this apparent limit in domain period have recently met with some success however. The first technique, referred to as controlled spontaneous backswitching, has been applied to bulk samples with a typical thickness of 500 $\mu$m, to generate periods of 4 $\mu$m,[8] and more recently 2.6 $\mu$m.[9] The second technique, applied to $MgO:LiNbO_3$ which has the benefit of improved resistance to photorefractive damage, utilized multiple short current pulses, generating a period of 2.2 $\mu$m and depth of 1.5 $\mu$m, which when used in conjunction with a waveguide geometry, has produced a high conversion efficiency.[10]

This last result is significant in that, for waveguide geometries at least, it is not necessary to achieve domain inversion to depths exceeding the guide depth itself. Many of the earlier reports on domain inversion applied to typical commercial material supplied as either 300 $\mu$m or 500 $\mu$m thick wafers. It is clearly harder to maintain high aspect ratio, short period, high-quality domain patterning over these large and (for waveguide geometries) unnecessarily large depths. In this letter, we discuss a method for achieving superficial, or surface, domain inversion that has been used to achieve periods of 1 $\mu$m, and that can be used, we believe, for achieving the periods of ∼0.3 $\mu$m required for waveguide implementation of backward wave parametric generation and tunable Bragg grating structures.

The technique for surface domain inversion is based on conventional electric-field ($E$-field) poling at room temperature. The procedure is as follows: One of the z faces of the crystal is covered with a photolithographically patterned photoresist layer with a thickness on the order of 1 $\mu$m in order to achieve the appropriate $E$-field contrast which is necessary for a spatially selective domain inversion. Both z faces are then covered with conductive gel electrodes, and a single high-voltage (HV) pulse is applied across the sample. The value of the HV varies with the thickness of the sample but the applied electric field must be on the order of 22 kV mm$^{-1}$.

For normal $E$-field poling, the established practice is to first calculate the charge, $Q$, corresponding to the patterned area intended for domain inversion. The formula used for this calculation is $Q = 2 \times A \times P_s$, where $Q$ is the calculated charge, $A$ is the area corresponding to the developed part of the photolithographic pattern (the area where the conductive liquid or gel is in contact with the crystal surface) and $P_s$ is the spontaneous polarization of lithium niobate (0.72 $\mu$C/mm$^2$). An additional external empirical factor ($EF$) is also usually taken into account, to correct for variations in supplier dependent material stoichiometry, precise values of thickness across the sample, and specific electrical characteristics of the poling supply itself. An $EF$ value exceeding unity is often used to achieve the desired high-quality peri-odic domain patterning, resulting in a calculated $Q$ value of $2\times A\times P_s\times EF$. The E-field is applied until the appropriate amount of charge according to the expression for $Q$ is detected. It is clear therefore, that the duration of the $E$-field application is a function of both the area to be poled and the $EF$ value.

[a]l:

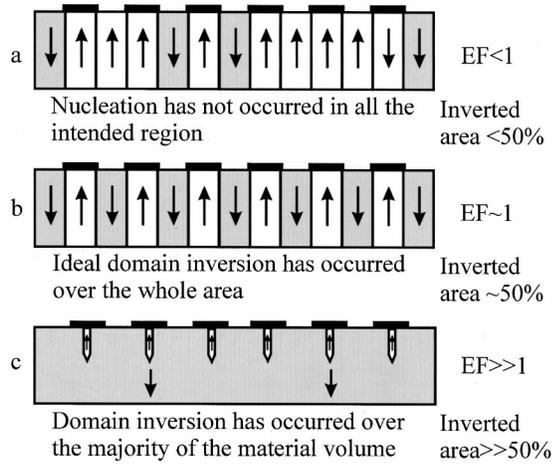

FIG. 1. Schematic of the poling process as a function of the empirical factor $EF$. (a) *underpoling* ($EF<1$), (b) normal poling ($EF\sim 1$), and (c) *overpoling* ($EF\gg 1$) used for fabrication of surface domain structures.

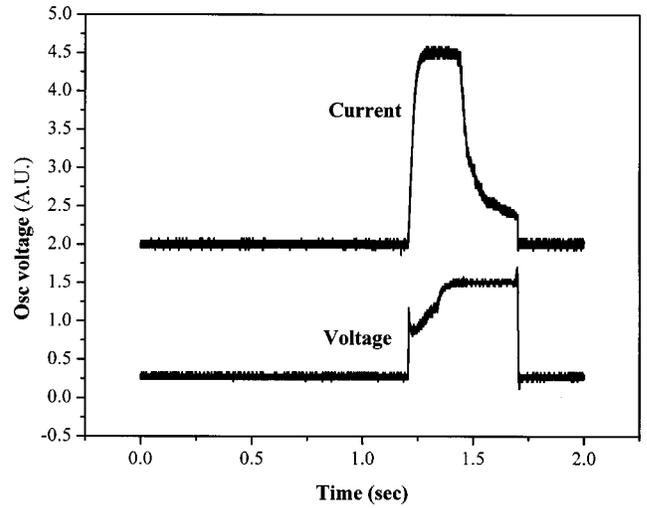

FIG. 2. Single-pulse poling signatures for current and voltage. Note that no backswitching is observed in this process.

This $EF$ thereby controls domain spreading within the crystal volume: Values of $EF<1$ lead to *underpoling*, whereby domains are inverted preferentially in areas where nucleation is easier, for example at the edges of the photoresist patterns or areas of increased surface roughness. If $EF\sim 1$, normal poling occurs which can be of good quality with 50/50 mark-to-space ratio and large scale uniformity for long period poled structures. Finally, if $EF$ is too large however, then the inverted domains, once nucleated, spread laterally, extending their volume more rapidly than required for an ideal 50/50 mark-to-space ratio grating. This case is referred to as *overpoling*. The three regimes, according to the domain spreading, are illustrated schematically in Fig. 1. Concentrating our attention on Fig. 1(c) which describes the state of the sample after poling using large values of $EF$, the schematic shows that small regions of material beneath the photoresist can remain in their original poled state. If overpoled, using values of $EF$ exceeding the theoretical value of $\sim 2$, then the sample appears almost uniformly poled when observed between crossed polarizers. Once etched with $HF/HNO_3$ acids, however, careful investigation reveals that some noninverted domain regions survive beneath the photoresist patterned surface, and that these can extend a few microns into the $-z$ crystal face. The technique which is described here relies on *overpoling* the sample which achieves the apparently undesirable effect of domain spreading and merging beneath the lithographically patterned photoresist layer. It is also able to create large scale uniform fine period surface inverted domain structures.

Using this technique, we have performed an initial parametric study of surface poling versus the value of $EF$ and imposed photoresist period. It should be noted that this technique will not work with other electrode materials such as directly deposited metals, as charge accumulation is thereby prohibited. We have used both conventional photolitho- graphic patterning for periods between 2.5 and 4.0 $\mu$m and also laser exposure through a phase mask. For the latter technique, final domain widths on the order of 0.5 $\mu$m and periods of 1 $\mu$m have been obtained. We have examined these surface domains for the former case, and found that they extend to depths of between 6 and 11 $\mu$m which is entirely compatible with waveguide depths and good overlap of guided modes.

For the larger range of periods studied, the $LiNbO_3$ $-z$ face was spin coated with a 1.2 $\mu$m thick photoresist and UV exposed through a periodic amplitude mask. After photoresist development, gel electrodes were applied to both the unpatterned $+z$ face and the patterned $-z$ surface. The samples were poled using a computer-controlled supply that dynamically varied the applied field in order to maintain a constant current, and the poling process terminated when a predefined charge $Q(=2\times A\times P_s\times EF)$ had passed through the crystal. A typical single-pulse poling curve is shown in Fig. 2, and illustrates the difference between this technique and that reported in Refs. 8 and 9. The applied E field is on the order of 22.1 kV/mm which is the appropriate value for domain reversal in lithium niobate. The high voltage is applied for a time duration that is proportional to the calculated charge value, hence, it will depend on the area to be poled as well as the $EF$ value. No backswitching occurs in our overpoling process, and we feel this represents a fundamentally simpler technique for achieving controlled small period surface domain inversion.

A variety of surface poling results can be obtained that depend on the value of $EF$ used. Figure 3 shows a typical scanning electron microscope (SEM) picture of a surface poled sample, obtained with an $EF$ value of 8. It is clearly seen following the $HF/HNO_3$ etching that the domains only exist in the near-surface region (shown here to a depth of $\sim 3$ $\mu$m). Other $EF$ values can and have been used but, to date, we have not performed a full parametric study of depth or uniformity as a function of the $EF$ value. The surface domain depth however is clearly an inverse function of the $EF$ value.

In Fig. 4, we show the results of measured domain depth as a function of the period of the imposed photolithographic pattern, for an *EF* value of 8. Although the variation of measured domain depth (taken for between 30 and 100 periods) is rather large, two clear points emerge. First, there is a minimum in the domain depth achieved, an obvious requirement for intended waveguide applications. Second, the mean depth is seen to scale approximately linearly with the period. For applications that require submicron periodicity, this is, again, a useful observation as the overlap between the guided modes and domain inverted regions is a prerequisite for efficient nonlinear interactions. Figure 4 shows two fits: One (dashed line) includes the point (0,0) as a further implicit data point. The close agreement between these two gradients further confirms the approximate linearity just stated.

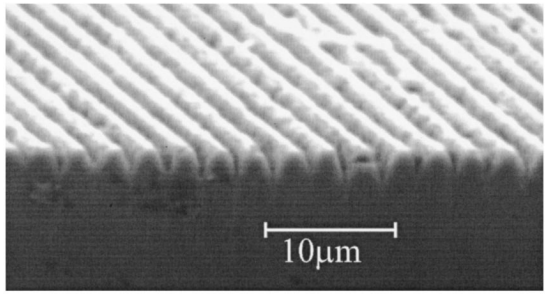

FIG. 3. SEM picture of surface domains revealed by HF/HNO$_3$ acid etching.

Finally, in Fig. 5, we show the details of a ~1 µm periodicity surface grating, fabricated using exposure of the photoresist via a phase mask. Following acid etching, sub-µm features are revealed that are on the order of 1 µm in depth. We believe that such interferometric exposure (via phase mask or two beam interferometry) holds much promise, as domain patterning down to periods on the order of 0.3 µm required for backward wave interactions at a wavelength of 1.5 µm should be readily achievable using exposure with near-UV laser irradiation.

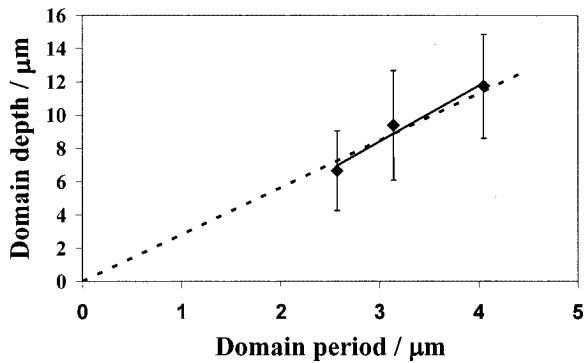

FIG. 4. Experimental measurements of domain depth vs domain period, as determined by optical microscopy, using an *EF* value of 8.

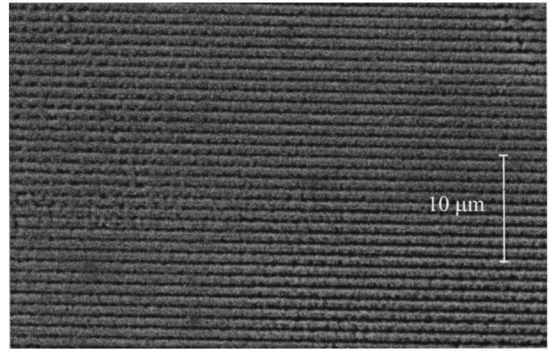

FIG. 5. SEM picture of 1 µm periodic surface domains written using a phase mask.

In summary, therefore, we have presented a single-step approach for achieving surface domain inversion to depths that are consistent with single-mode waveguides in LiNbO$_3$. The overpoling technique is simple to implement, and appears to work down to periodicities of at least 1 µm. Further work is in progress to examine the optimum choice for the *EF* value used, and to fabricate first-order gratings in waveguide materials, with the required periodicities of ~2 µm.

The authors are pleased to acknowledge support from the Engineering and Physical Sciences Research Council (EPSRC) for research funding, under Grant No. GR/R47295, and thank Peter G. R. Smith from the ORC, University of Southampton, for discussions.